# Factors Influencing the Performance of Students in Software Automated Test Tools Course

Susmita Haldar
*School of Information Technology*
*Fanshawe College*
London, Canada
shaldar@fanshawec.ca

Mary Pierce
*Faculty of Business,*
*Info Technology and Pt Studies*
*Fanshawe College*
London, Canada
mpierce@fanshawec.ca

Luiz Fernando Capretz
*Department of Electrical and Computer Engineering*
*Western University*
London, Canada
lcapretz@uwo.ca

*Abstract*—Formal software testing education is important for building efficient QA professionals. Various aspects of quality assurance approaches are usually covered in courses for training software testing students. Automated Test Tools is one of the core courses in the software testing post-graduate curriculum due to the high demand for automated testers in the workforce. It is important to understand which factors are affecting student performance in the automated testing course to be able to assist the students early on based on their needs. Various metrics that are considered for predicting student performance in this testing course are student engagement, grades on individual deliverables, and prerequisite courses. This study identifies the impact of assessing students based on individual vs. group activities, theoretical vs. practical components, and the effect of having taken prerequisite courses in their final grade. To carry out this research, student data was collected from the automated test tools course of a community college-based postgraduate certificate program in software testing. The dataset contained student records from the years 2021 to 2022 and consisted of information from five different semesters. Various machine learning algorithms were applied to develop an effective model for predicting students' performance in the automated software testing tools course, and finally, important features affecting the students' performance were identified. The predictive performance model of the automated test tools course that was developed by applying the logistic regression technique, showed the best performance, with an accuracy score of 90%.

*Index Terms*—Software Testing Education, Automated Testing, Selenium, Quality Assurance, Student Engagement

## I. Introduction

Software testing is critical to delivering software with fewer defects and in turn with good quality. Industries rely on their QA team to verify that the software has been fully tested. Manual testing can be very labor-intensive, monotonous, and error-prone. As a result, to ensure the delivery of quality products, workplaces focus on automating manual test scripts to increase test coverage. Automated testing requires specialized skills in addition to having the mindset of a manual tester, which includes an understanding of coding concepts, knowledge of the IT domain, and a basic understanding of software testing.

However, a software testing career is a less favorable choice within the software engineering domain [1]. Capretz et al. [2] conducted a study on the software testing profession, where they identified the factors that motivate or demotivate software testing professionals to sustain their software testing career and found that only 25% of the selected respondents from Canada were motivated to take software testing as a career option. Therefore, when choosing from among the few experienced QA professionals, companies want to ensure that these individuals can recommend the required tools for automating the manual testing effort and develop and execute automated test scripts to ensure higher test coverage.

To bridge this gap, the post-graduate certificate program in software testing offers a dedicated course in automated test tools which is worth four credits and can be taken in the second level of a two-level postgraduate program. Students are only permitted to take this course after completing prerequisite courses containing foundation-level concepts in software testing and JAVA programming language. The college needs to be careful to choose only the prerequisite courses that are essential for students' learning. Keeping unnecessary prerequisites in a course can demotivate students who are ready to take on challenges of advanced concepts. It can also affect the student retention strategy as the students will be restricted from taking this course if they have failed the prerequisite courses in level one. On the other hand, exceptional students may want to take on a challenge by taking an advanced course in their program to keep them motivated without waiting for completion of the prerequisite courses.

If only theoretical concepts are taught in a software testing course, the student's knowledge may be far from the real-world expectations [3]. At the same time, teaching hands-on problems only may not cultivate the critical thinking strategies required for developing an effective automated software testing (AST) solution. In addition, instructors need to understand whether practical exercises are affecting the students' final grades compared to assessments on theoretical quizzes or assignments to customize the course delivery to meet the needs of the students.

Student engagement factors such as the number of times the student logged into the learning management system, and the number of video recording contents completed by the student can assist the instructor and the academic advisor in guiding the students to focus on studying. The number of assignment submissions, and quiz submissions demonstrates whether the student missed tests or assignments because of being ignorant about them. This research will also assess the contribution of student engagement to students' final grades. A minimum C grade is desirable as a cumulative GPA lower than a C grade can impact their academic standing status.

Teaching students AST concepts, techniques, and theories is a challenging task, especially where the focus is also on vocational learning outcomes, and essential employability skills in addition to course learning outcomes. Software engineering students often do not gain the essential skills and knowledge they need to succeed in the IT industry [4]. Although most Computer Science programs offer some testing concepts in their programs, they often fail to remain aligned with the realities of the IT industry [5].

Applying machine learning algorithms, this paper will investigate whether students' performance in automated software testing tools course can be predicted based on their assessment from milestone deliverables, their engagement in class, prerequisite courses, etc.

The rest of this paper is organized as follows. Section II provides background on Software Testing education, highlighting automated software testing education. Section III describes how the AST course has been structured, and provides the methodology for developing the model for predicting students' performance on the AST tools course. Section IV presents the results of this study. Finally, section V provides analysis and discussion. This is followed by the paper's conclusions and suggestions for future work in section VII.

## II. BACKGROUND AND LITERATURE SURVEY

### A. Software Testing Education

There has been an emphasis on considering software testing in the Software Engineering or Computer Science curriculum. Garousi et al. [6] conducted a study on the state of software testing education in Canadian and American universities and identified the strengths and areas for improvement. As part of their recommendations, they encouraged a systematic software testing curriculum. They also remarked that Computer Science and software engineering graduates need to be able to perform testing and quality assurance tasks on the software they produce after graduation. They found that even some top university programs were not offering any courses in the software testing domain.

### B. Automated Software Testing Education

Teaching only certain testing techniques may not provide the skillset to apply automated testing skills in the industry. As a result, teaching automated test tools as a separate course can give proper emphasis to the professors to teach students how to utilize various AST tools effectively instead of focusing on manual testing only. Barrett et al. [7] surveyed 25 courses offered by 14 universities in Sweden. From their analysis of the basic curriculum, it appears that utilization of AST tools was incorporated with other components in the majority of the courses, but it appears that automated testing was not offered as a separate course in most of these selected universities.

### C. Student Performance Prediction

Student performance prediction is an emerging research area, but this approach has not been applied to software testing students, even though this control group can demonstrate different characteristics than others. For instance, in previous studies, one of the challenges shown with teaching software testing was keeping students' motivation in testing courses alive compared to their motivation in other subjects [8], [9]. Different researchers applied various techniques for teaching software testing, such as using a game-based approach [10], [11], utilizing free and open-source software [12], and using Selenium [13] for creating automated test solutions [14], etc.

Mohammaed et al. [15] applied the Apriori rule-based algorithm to identify the relationship between student engagement and student performance. They found that highly engaged students were performing well in the courses.

Burman et al. [16] applied a multi-class Support Vector Machine (SVM) classification model to classify the learners in the categories of high, average, and low, according to their academic scores. They then applied linear kernel and radial basis kernel. RBF produced better results than the linear kernel for predicting students' performance and the Radial Basis Function kernel gave more accurate results than the Linear Kernel, which showed an accuracy of approximately 91%.

Bhutto et al. [17] applied logistic regression and SVM for student performance prediction. They obtained 73% and 70% accuracy using logistic regression and SVM respectively. Bujang et al. [18] used Decision Tree, SVM, Naive Bayes, KNN, Logistic Regression, and Random Forest(RF) on the course grade datasets of real students, incorporating 1282 records. They proposed a multiclass prediction model to reduce the over-fitting and miss-classification results caused by imbalanced multi-classification, based on the Synthetic Minority Oversampling Technique (SMOTE). Their proposed model, integrated with RF, gave the highest F1 score of 99.5%.

Khan et al. [19] used data mining techniques based on LMS activity logs and applied a rule-based algorithm for predicting student performance. They found that there is a considerable correlation between student performance and several different factors, such as resource views, activity gaps, grades from the previous semester, grades from prerequisite courses, and evaluations of first-term tests. In this research, we will also look at student engagement metrics to evaluate student performance in the automated software testing tool course. Professors and academic advisors can use this study to spot students who need extra assistance so they can intervene.

Shi et al. [20] analyzed the characteristics of college students' learning behaviors and explored the predictive learning effect by constructing a machine learning model of learning effect based on information literacy learning behavior characteristics. Out of several algorithms, they attempted, the RF model showed the best performance with a value accuracy of 92.50%, Precision of 84.56%, Recall of 94.81%, F1-Score 89.39%, and Kappa coefficient of 0.859%.

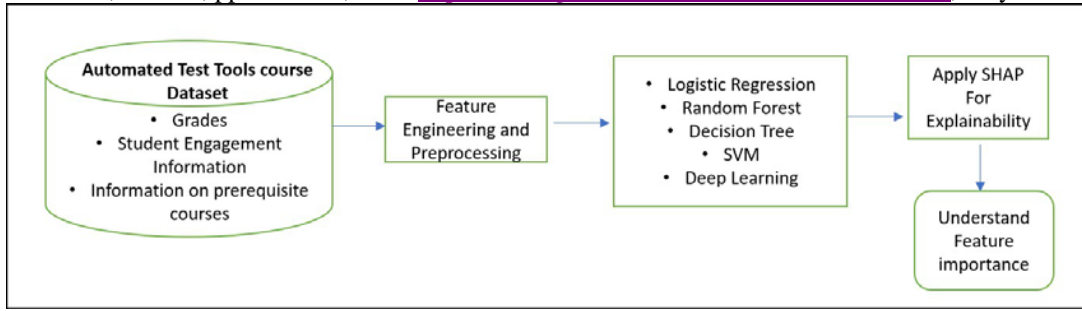

Fig. 1. Methodology applied for predicting the final performance of students of Automated Test Tools course

Jayasundara et al. [21] conducted a study on building an explainable boosting student performance model with a dataset from university entrance exam performance. They considered the interpretability of the model by demographic information such as gender, caste, parent's education, and previous educational background. The accuracy of the developed model had a maximum F1 score of 77% on predicting average grades for students while the other categories such as good, bad, and excellent grade detection rates were relatively low.

This research will contribute to the software testing education community by helping us to understand which factors are important for a course in automated testing.

## III. METHODOLOGY

The methodology of the program has been demonstrated in Fig. 1. The first step was to collect the data and apply feature engineering and pre-processing. Selected machine learning algorithms were applied for the student performance model development. The performance of the developed models was evaluated according to the selected criteria. On the best-performing model, the SHAP technique was applied to understand the feature importance of the model. The details are described below.

### A. Data Collection

The development of the model was initiated with data collection from the one-year postgraduate software testing certificate program available under the Department of Information Technology at Fanshawe College of Applied Arts and Technology located in London Ontario. Automated Test Tools is a second-level course among a total of 11 courses available in the program. The data was collected from various sources, especially from the database where grading information is stored. The dataset was anonymized due to having protected information. Student-related information, such as date of birth, actual student ID, etc., was removed from the dataset. The dataset consists of students from five different semesters, and the duration represents the classes taken during the pandemic when the classes were taken online. The total sample size of the dataset is 223.

This work utilized established machine learning algorithms applicable to classification problems with mid-sized datasets.

The expected output was split into five different categories which are D, B, C, A, and A+.

### B. Feature Selection

The evaluation strategy of the Automated Test Tools course has been illustrated in Table I.

TABLE I
PRIMARY DISTRIBUTION OF GRADE COMPONENTS

| Grade Item | Submission Type | Type of assignment | Worth |
|---|---|---|---|
| Research Projects | Group submission | Practical focused | 45% |
| In-class exercises | Individual submission | Practical focused | 15% |
| Quizzes | Individual submission | Theory focused | 40% |

The research projects are worth 45% of the final grade and are conducted over the semester through group collaboration. Forty percent of the overall grade is taken from class tests, and the remaining 15% is dedicated to in-class exercises. The research projects are divided into three projects with the same teams of 3-4 students working together on the projects, for the full term. This research work evaluation is often split between 40% on projects which require teamwork, and 5% on individual class attendance based on the professor's interest. The research projects consist of three group projects reflecting the development of automated test solutions. The projects involve software testing-related activities starting from gathering requirements, designing, and implementing an automated regression suite repository using an open-source tool called Selenium IDE [22], moving forward with automated test data generation using available open-source tools, and finally building automated test solution using Selenium WebDriver, JUnit [23] and TestNg [24] and deploying the code in DevOps tool Jenkins [25].

Table II shows a sample rubric of how Project 1 worth around 13.33% is being evaluated. The following are the mapped course learning outcomes for this project:

- Discuss what it means to implement an AST solution.
- Describe the benefits of implementing AST.
- Design an automated testing strategy for a baseline software application.

As the description suggests, the research projects require the technical knowledge to explore various options for implementing an effective AST solution. The group projects give the students exposure to the practical aspects of software automation through collaboration, interaction, and brainstorming with team members, and the successful implementation of the projects.

TABLE II
RUBRICS FOR PROJECT 1 WHICH IS A GROUP PROJECT FOR IMPLEMENTING AUTOMATED TESTING SOLUTION USING SELENIUM IDE

| Criteria | Excellent | Average | Poor |
| --- | --- | --- | --- |
| SUT Selection | The selected website was an excellent candidate because of the opportunity to test various features, and the justification of selection was provided in detail. | The selected website was considered average as it offers some features for testing, but with a limited scope. The document lacks sufficient reasoning for its selection. | The selected website was a poor candidate for this AST project due to not having enough testable features. The document did not provide proper justification. |
| Requirements Traceability Matrix | The RTM was very thorough and demonstrated how requirements are linked to the test cases. | Some requirements are mapped to test cases. | The creation of RTM was done poorly. |
| Selection of Testcases | Efforts in creating the testcases is very prominent, and a sufficient number of testcases were attempted. | Testcase selection is showing an average effort. | Not enough test cases were automated. |
| Test Scripts Document | Actual results for all the scenarios - completed and traceable. | Actual results for major scenarios - completed and traceable. | Actual results for scenarios - Incomplete or not traceable. |
| Automated Scripts | Automated tools were used comprehensively, and results were generated correctly. | Automated tools were used okay, and most of the results were generated correctly. | Automated tool was insufficiently used, resulting in incomplete results. |
| Documentation | Carefully followed all instructions, adhering to document format/presentation, spelling, and grammar standards, including providing sufficient screenshots. | Followed most provided instructions, document format/presentation, spelling, and grammar including providing some screenshots. | Did not follow instructions, and the document did not contain all the requested contents. |
| Source Code | Source code is submitted and accompanied by thorough documentation and clear explanations, ensuring a comprehensive understanding of its functionality | Source code was submitted, but the source code is somewhat comprehensive. | Source code was not submitted or the submitted source code is of poor quality, and not comprehensive. |
| Critical thinking & problem solving skills | The implemented solution demonstrates excellent critical thinking and problem-solving skills. | The implemented solution demonstrates somewhat critical thinking and problem-solving skills. | The implemented solution demonstrates limited critical thinking and problem-solving skills. |

Grade items on the test, worth 40% of the total mark, assess students on their understanding of theoretical concepts of software testing and are equally distributed among four different tests. The tests are individual deliverables and usually have multiple-choice questions along with a few scenario-based questions. The in-class exercises, which entail 15% of the final grade, let the students compare manual and automated efforts, implement automated solutions using an open-source tool called Katalon Recorder [26], and finally write test cases using Selenium WebDriver [13].

As part of data preparation and preprocessing activities, the percentage of research scores, test scores, and in-class exercise scores were calculated for each student, and the percentage scores were considered. Next, the system also contains information about how many times the student accessed the course page, how many of the available content or videos have been reviewed by the students, the number of assignments submitted in the course, and the number of quizzes completed. These are indicators of student engagement. For instance, the number of available content modules students are completing, along with their positive motivation towards completing the available recorded video lectures, may demonstrate that students are engaged in this course.

Afterward, the student's grades in the two prerequisite courses were added to the dataset. As discussed in previous sections, an interest of this study was to assess whether the grade in the prerequisite courses assisted with the student's performance prediction. The students' past failure history in the pre-requisite courses was calculated and considered as a separate input feature for developing the predictive model. A value of 1 indicates the student needed to retake the prerequisite course, and 0 indicates the student was able to pass the course with a single attempt.

The prerequisite course 'Test Methodology' gives students a background in quality assurance methodologies, including black-box, white-box, grey-box, unit, and other testing methods. The other prerequisite course is 'Coding for Tests'. This course examines the practices and procedures related to creating and debugging software. This course also prepares the student to write code, using a procedural approach initially, and then migrating to an object-oriented approach. The final feature list has been demonstrated in Table III. This table shows the 12 features selected as input for developing the model that would predict the letter-grade-category field.

*C. Data Pre-processing*

The data cleaning step verified the dataset for any null or redundant values. The data was split into 70% for training and 30% for testing to ensure that both training and testing datasets contained representation from each selected grade category of D, C, B, A, and A+. The training and testing data was normalized using RobustScaler [27], a functionality available from the scikit-learn library of Python, to scale features using statistics that are robust to outliers. RobustScaler removes the median and scales the data according to the quartile range.

TABLE III
AUTOMATED TEST TOOL - STUDENT PERFORMANCE DATA

| Feature Name | Type | Description |
|---|---|---|
| ContentCompleted | Numeric | The total number of views of the contents that were uploaded to the course home page. |
| QuizCompleted | Numeric | Total number of quizzes completed. Students are required to write four quizzes. |
| #OfAssignmentSubmissions | Numeric | Total number of assignments submitted. |
| CourseAccess | Numeric | Number of times the student has accessed the course home page. |
| CodingforTest_score | Numeric | Students' grade in the prerequisite course of "Coding for Test". This course gives exposure to writing code |
| TestMethodology_score | Numeric | Students grade in the prerequisite course of "Testing Methodology Course". This course introduces the student to a myriad of QA methodologies, including black-box, white-box, grey-box, unit, and other testing methods. |
| Retook TestMethodology | Numeric | A value of 0 if the student never failed prerequisite "TestMethodology course", and 1 otherwise. |
| Retook CodingForTest | Numeric | A value of 0 if the student never failed prerequisite "Prerequisite Coding for Test course", and 1 otherwise. |
| TestScore | Numeric | Total score in the number of tests taken in the course. This is worth 40% of the final grade. |
| ResearchScore | Numeric | Total score on research projects. The research projects are done in group settings. |
| InClassExerciseScore | Numeric | Total score on in-class exercises done as independent assignments, and worth 15% of the final grade. |
| MidtermScore | Numeric | The total score is a summation of all the assignments done before the midterm reporting deadline. This may reflect the score in tests, research, and in-class exercises that are completed before the midterm due date. |

### D. Applied Machine Learning Algorithms

As our dependent variable consists of different categories, including A+, A, B, C, and D, we have applied machine learning algorithms that apply to common classifiers that work for multi-classification problems. The multi-classification problem has more than two possible classes or variables.

In literature, Alamri et al. [28] surveyed various machine learning algorithms that have been employed in explainable student performance studies, and it appears that the Decision Tree Algorithm and rule-based algorithms have been successful in developing the models for multi-class classification problems. In a similar vein, this study also considered the explainability of the model in terms of feature selection and investigated the effectiveness with simpler machine learning techniques of Logistic Regression [29], SVMs [30], and Random Forest Model, and then moved to the complex technique of DFFN (Deep Feed-Forward Neural Network). Our study compared these classification algorithms in their ability to detect student performance.

**Logistic Regression** predicts the probability that an instance belongs to a particular class [31]. The target variable has five different ordinal grade categories. Multiclass Logistic regression is extended from a binary classification using one versus all or one versus rest method. One-vs-rest (OVR) classifier involves training a single classifier per class, with the samples of that class as positive samples and all other samples as negatives.

**Support Vector Machine** SVM is a supervised learning technique that aims to classify the data. It uses a hyperplane for dividing the dataset into classes with a gap as wide as possible known as a margin. For considering this multiclassification problem with predicting the selected five types of grades, the problem was broken down into multiple binary classification problems. Similar to logistic regression, the one-vs-rest technique was used. [32]

**Decision Tree Classifier** can be applied in both binary and multiclass classification problems. This algorithm is useful for relatively small datasets that have a simple underlying structure, and this model is easily interpretable [33].

**Random Forest** [34] is an ensemble classification technique. This algorithm constructs many decision trees like a forest with random attribute values.

**Deep Feed-Forward Neural Network** [35] is the simplest type of artificial neural network that has various applications in machine learning. DFNN is a suitable candidate for multiclass classification and works with larger datasets.

### E. Evaluation Metrics

We have used the traditional evaluation methodology for comparing the effectiveness of the selected multi-class classification models which are **Accuracy**, **Precision**, **Recall**, **F1-Score**, **ROC_AUC Score**.

**Accuracy** It is the ratio of correctly predicted instances to the total instances.

$$Accuracy = \frac{Number of Correct Predictions}{Total Number of Predictions} \quad (1)$$

Although accuracy is considered an important metric for measuring the performance of a problem, additional measures are required to ensure the result is not biased toward predicting a single category of students. To ensure the reliability of the model, all different types of grades should be predicted. The following evaluation metrics were considered in addition to accuracy. **Precision** is the ratio of true positive predictions to the total positive predictions. It focuses on the accuracy of positive predictions. **Precision** has been defined by the following equation:

$$Precision = \frac{True Positives}{True Positives + False Positives} \quad (2)$$

**Recall** is the ratio of true positive predictions to the total actual positive instances. It measures the ability of the model to capture all positive instances.

$$Recall = \frac{True Positives}{True Positives + False Negatives} \quad (3)$$

**F1 score** is the harmonic mean of precision and recall. It provides a balanced measure between precision and recall.

$$F1 Score = 2 \times \frac{Precision + Recall}{Precision \times Recall} \quad (4)$$

**AUC ROC curve** is a graphical representation of the model's ability to distinguish between positive and negative instances. The AUC (Area Under the Curve) summarizes the ROC curve into a single value. ROC_AUC score is a single number that summarizes the classifier's performance across all possible classification thresholds.

*F. Feature Importance*

After the best model is selected, the features that contributed to the model's development will be analyzed based on a model-agnostic approach. To rely on the model interpretation, SHapley Additive exPlanations (SHAP), a game-based theory approach has been used to explain the machine learning model [36]. Finding a consistent and impartial explanation of how each characteristic affects the model's prediction is commonly accomplished through the use of SHAP values. The significance of each feature in a model is assigned by SHAP values, which are based on game theory. How strong the influence is shown by its magnitude.

IV. RESULTS

Table IV demonstrates the parameters that were selected for each of the selected algorithms after applying hyperparameter tuning using the grid-search algorithm.

TABLE IV
HYPERPARAMETERS UTILIZED IN THE MACHINE LEARNING MODELS.

| Model Name | Best Parameter |
|---|---|
| Logistic Regression | C: 120, class_weight: balanced, max_iter: 1000, multi_class: multinomial, penalty: l2, solver: lbfgs. |
| SVM | estimator_C: 23, estimator_gamma: 0.05, estimator_kernel: rbf. |
| Random Forest | max_depth: 30, min_samples_leaf: 1, min_samples_split: 5, n_estimators: 100. |
| Decision Tree | criterion: entropy, max_depth: None, min_samples_leaf: 1, min_samples_split: 2, splitter: best. |
| Deep Neural Network | batch_size: 40, epochs: 700, model_activation: relu, model_dropout_rate: 0.2, model_hidden_units: (45, 36, 27, 18), model_optimizer: adam. |

Table V summarizes the result of applying the selected algorithms. Logistic Regression outperformed all other models in multiple evaluation metrics. The logistic regression model had the highest value for accuracy, precision, recall, and F1_Score, and the second highest score for ROC_AUC score. The next best-performing model was the Deep Feed Forward Neural Network with an accuracy score of 85% compared to 90% from Logistic Regression, a precision value of 83% which is less than a precision value of 89% from Logistic Regression, and an equal value with Random Forest classifier. The Recall and F1 Score of the DFNN network are in second position compared to other models presented in this study.

The AUC_ROC score of the DFNN model was the highest among all models, which means it detected the prediction from each of the categories. The Random Forest algorithm had a score of about 80% in all the evaluation metrics but had a slightly lower score than DFFNN and lower than Logistic Regression in all categories of evaluation metrics. SVM and Decision Tree classifiers were the poorest performers compared to Logistic Regression, with a score of between 70% to 79% in all categories except the AUC_ROC score.

We, therefore, considered the Logistic Regression model the best-performing model because although the ROC_AUC score is slightly lower than the DFNN model with a value of 98% compared to 99%, all other evaluation metrics achieved higher values for the Logistic Regression model.

For verifying the feature importance, the SHAP method has been applied to the best-performing Logistic model. The generated summary plot of SHAP values has been shown in Fig. 2. The X-axis of the bar graph shows the mean absolute SHAP values while the y-axis demonstrates the feature contributions from highest to lowest ranking. The feature importance is calculated based on the aggregated values of student records from five different semesters. The legend shows the instances taken from the five different selected student grades which are A+, A, B, C, and D. By examining these colored segments, we can understand how each feature influences the model's predictions for each category. This detailed visualization allows interpretation of feature importance, shedding light on the specific role of each feature in differentiating between each grade type, and we can thus understand how each feature influences the model's predictions.

This bar graph shows that the highest contributor to the final grade is the research score, followed by the test score, and in-class exercise score. As these scores are direct components of the final grade, these features contribute the most to the model aligning with the expectation.

Although the research score has the highest impact on the model contribution, observing at a more granular level shows that for students with A+ grades, the testing score was more important than the research score. On the other hand, for students with a D grade, the highest impact came from research scores.

The next important feature shown in the bar graph of this student performance model is the midterm score. This feature is not a direct component of the final grade but is an indication of a student's progress in the middle of the term. Satisfactory or unsatisfactory status is provided based on the accumulated score from other components until the midterm grade submission time. The midterm grade is usually reported between the 6th to 7th week of a 15-week term. This result illustrates how the student's performance at the halfway point of the term contributes to predicting the student's final grade. The result shows the midterm score is one of the important, but not the most important contributors. This could be due to the first several weeks of the classes delivering fundamental concepts. More challenging concepts and assignments are available in the second part of the course due to the requirement of understanding advanced concepts.

TABLE V
RESULT OF RUNNING THE CLASSIFIERS

| Model Name | Accuracy | Precision | Recall | F1 Score | AUC |
|---|---|---|---|---|---|
| Logistic Regression | 90% | 89% | 90% | 89% | 98% |
| SVM | 75% | 70% | 75% | 72% | 97% |
| Decision Tree Classifier | 78% | 77% | 78% | 77% | 77% |
| Random Forest Classifier | 84% | 83% | 84% | 82% | 97% |
| Deep Feed Forward Neural Network | 85% | 83% | 85% | 84% | 99% |

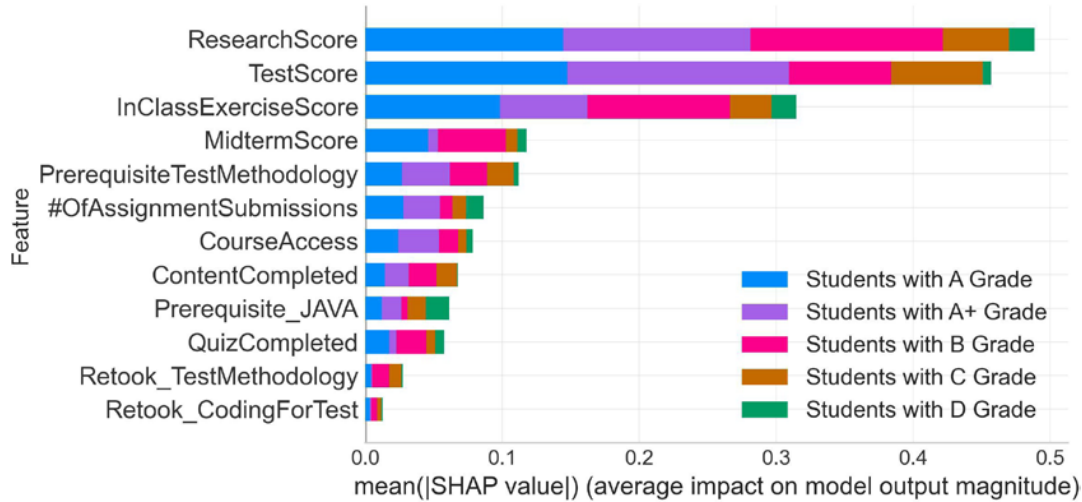

Fig. 2. Feature Importance from SHAP model

The next attribute of the student's grades prediction model is their prerequisite test methodology. This is not a direct component of the final grade. However, as this course provides a foundation-level knowledge of testing, the student's competency in this course was measured using their grades and appears to affect students' outcomes in the AST tool course.

However, the score of the prerequisite course is not the primary contributor as the students can still manage to do well in research work through group studies, independent work toward tests, and by being attentive during in-class exercises. At the same time, this prerequisite course cannot be waived from the course since the SHAP model demonstrates one of the important indicators of this model. Also, not having enough background in this prerequisite course may set the students to achieve the learning objective of this course.

The next three features reflecting student engagement are the number of assignment submissions, course access, and content completed in order of feature contributions to this model respectively. If the students don't submit all their assignments as expected, that means they were not regular in class. Hence, the strong students will try to abide by the due date and submit their assignments accordingly. However, the students who skip their assignments entirely may be indicating that they are not serious about their grades. The course access attribute demonstrates the number of times the student accessed the course home page. This set of students' records reflects when the students were moved to virtual classes. Most of the students were attending the synchronous lectures, and being up to date with the lectures can assist with their overall performance in the course. After the synchronous lectures, professors used to post the video lectures and other materials for students to review. This next feature shows that students' completion or review of the content also helped with their overall performance, and vice versa.

Another component that was used for developing this model was the grade in their prerequisite coding for tests course where the JAVA programming language was introduced. The Coding for Test course did impact the students' overall performance prediction, but except for students who received D grades, it appears the priority is on the student's background in the prerequisite testing methodology course compared to the Coding for Test course. In this course, through group research projects, students review and practice skills learned in the coding for test course. The score on the research project is important as the weight of this component in the final grade is 45%. The SHAP model also conforms to this by showing the research score as the strongest contributor to this course. As the students need to do coding in the automated test tools course, they will certainly suffer in the automated testing course without prior knowledge of coding.

QuizCompleted is an attribute from the student engagement category. Although this factor impacts the outcome of the student performance prediction model, it is not one of the major contributors. The reason could be that the majority of the students don't skip their quizzes as each of the tests is worth 10% of their total score. In exceptional scenarios, they request a rescheduling of the exam.

The last two factors affecting this model are the students who had a history of failure in the prerequisite courses of test methodology and coding for tests respectively. As the students cannot take the automated testing course without passing these two prerequisite courses, the previous failure history on these courses does not significantly impact this student performance model. The main factors are whether the students work hard towards their group projects, study for their tests, and participate effectively in their in-class exercises, followed by their previous background in testing methodology courses and engagement towards their course material.

## V. ANALYSIS AND DISCUSSION

The performance of the student performance prediction model is comparable to other previous studies. For instance, the student performance model developed by Burman et al. [16] showed that the best-performing model had an accuracy of 91% whereas our implementation using the logistic regression model showed an accuracy of 90%. However, the other evaluation criteria such as precision, recall, F1-Score, and ROC_AUC score were not measured in Burman et al.'s study. Since the result contains multiple grade categories, we wanted to ensure all different grade categories have been represented properly through evaluation criteria in addition to accuracy. The precision, recall, F1-Score, and ROC_AUC scores of the developed models are 89%, 90%, 89%, and 98% respectively. These high scores bring confidence in the developed model.

Bujang et al. [18] achieved an F1_score of 99.5% compared to the score of 89% from our best-performing model. In their study, the oversampling strategy was used to generate more data for each different great category. Without applying oversampling techniques, we were able to achieve closer to the 90% range for the F1 measure. This study was able to predict how the prerequisite courses and course engagements can also impact the student's academic performance compared to other studies.

The SHAP explanation from the obtained model shows that students can learn effectively when they work in a group setting by viewing the highest ranking in research projects. An independent study on gaining theoretical background is also important because it can lead to doing well in research work. Student engagement through in-class activities is important for students' success in their courses. The practical components applied during in-class exercises use open-source tools in addition to in-house applications when validating the selected software being tested. This combined approach of theory with hands-on experience is developing effective QA professionals to be ready for the job market.

## VI. THREATS TO VALIDITY

This work was based on the accumulation of data over five semesters during the pandemic period. The ability to provide more data including from the post-pandemic period may help us to generalize the findings. In addition, depending on the program, and student enrollment requirements, the prerequisite requirements can vary. This study was based on a postgraduate certificate program where students usually have an IT background with a diploma in the business or IT domain with or without a combination of work experience. This finding may vary slightly for students taking software testing courses without any Computer Science background.

## VII. CONCLUSION

This study investigated the performance of the student performance prediction model of the software automated testing tool from the software testing post-graduate certificate program. Software performance prediction is an emerging research area, but this has not been applied to assessing academic performance in the software quality assurance domain. Software testing students show different personality traits and often have less motivation toward a software testing career compared to other software engineering students.

Five different machine learning algorithms were attempted, and logistic regression performed best out of these models, with scores close to the 90% range in accuracy, precision, recall, F1 score, and ROC AUC score. This study has found that a student's regular submissions of deliverables, engagement in class, and background in the testing foundation and programming concepts–along with dedication to the group and independent work with a hybrid knowledge in theory and practical concepts on developing effective software testing solutions—helps with performance prediction.

This study can be extended by investigating whether other prerequisite courses should be added to this course. This could be achieved by assessing the grades of other level-one courses in the program. Few other courses in level one cover a curriculum consisting of soft skills such as communications, academic integrity, and applied project management. Our future software testing professionals will require both soft skills and technical skills to enable them to be ready for the real world.

Finally, this study was able to develop an effective student performance model for software-automated test tools courses with the capability of explaining important factors for predicting student performance.

## ACKNOWLEDGMENT

The authors would like to acknowledge Dr. Dev. Sainani, the associate dean of the School of Information Technology of Fanshawe College for supporting this research work. The authors would like to extend their thanks to Learning Systems Services, Robert R Downie, the Institutional Research Department, and the REB board of Fanshawe College for assisting with the data collection process.

## REFERENCES

[1] R. d. S. Santos, L. F. Capretz, C. V. C. de Magalhães, and R. Souza, "Myths and Facts about a Career in Software Testing: The Perspectives of Students and Practitioners," in *2023 IEEE 35th International Conference on Software Engineering Education and Training (CSEE&T)*, Aug. 2023, pp. 120–120, iSSN: 2377-570X. [Online]. Available: https://ieeexplore.ieee.org/document/10229341

[2] L. F. Capretz, P. Waychal, J. Jia, D. Varona, and Y. Lizama, "Studies on the Software Testing Profession," in *2019 IEEE/ACM 41st International Conference on Software Engineering: Companion Proceedings (ICSE-Companion)*, May 2019, pp. 262–263, iSSN: 2574-1934. [Online]. Available: https://ieeexplore.ieee.org/document/8802688

[3] S. M. Melo, V. X. S. Moreira, L. N. Paschoal, and S. R. S. Souza, "Testing Education: A Survey on a Global Scale," in *Proceedings of the XXXIV Brazilian Symposium on Software Engineering*, ser. SBES '20. New York, NY, USA: Association for Computing Machinery, Dec. 2020, pp. 554–563. [Online]. Available: https://dl.acm.org/doi/10.1145/3422392.3422483

[4] D. Oguz and K. Oguz, "Perspectives on the gap between the software industry and the software engineering education," *IEEE Access*, vol. 7, pp. 117 527–117 543, 2019.

[5] M. Aniche, F. Hermans, and A. van Deursen, "Pragmatic software testing education," in *Proceedings of the 50th ACM Technical Symposium on Computer Science Education*, ser. SIGCSE '19. New York, NY, USA: Association for Computing Machinery, 2019, p. 414–420. [Online]. Available: https://doi.org/10.1145/3287324.3287461

[6] V. Garousi and A. Mathur, "Current State of the Software Testing Education in North American Academia and Some Recommendations for the New Educators," in *2010 23rd IEEE Conference on Software Engineering Education and Training*, Mar. 2010, pp. 89–96, iSSN: 2377-570X. [Online]. Available: https://ieeexplore.ieee.org/abstract/document/5463575

[7] A. A. Barrett, E. Paul Enoiu, and W. Afzal, "On the Current State of Academic Software Testing Education in Sweden," in *2023 IEEE International Conference on Software Testing, Verification and Validation Workshops (ICSTW)*, Apr. 2023, pp. 397–404, iSSN: 2159-4848. [Online]. Available: https://ieeexplore.ieee.org/document/10132264

[8] D. Towey and T. Y. Chen, "Teaching software testing skills: Metamorphic testing as vehicle for creativity and effectiveness in software testing," in *2015 IEEE International Conference on Teaching, Assessment, and Learning for Engineering (TALE)*, Dec. 2015, pp. 161–162. [Online]. Available: https://ieeexplore.ieee.org/abstract/document/7386036

[9] G. Fraser, A. Gambi, M. Kreis, and J. M. Rojas, "Gamifying a software testing course with code defenders," in *Proceedings of the 50th ACM Technical Symposium on Computer Science Education*, 2019, pp. 571–577.

[10] B. S. Clegg, J. M. Rojas, and G. Fraser, "Teaching Software Testing Concepts Using a Mutation Testing Game," in *2017 IEEE/ACM 39th International Conference on Software Engineering: Software Engineering Education and Training Track (ICSE-SEET)*, May 2017, pp. 33–36. [Online]. Available: https://ieeexplore.ieee.org/abstract/document/7964327

[11] J. Andrews, "Killer app: A eurogame about software quality," in *2013 26th International Conference on Software Engineering Education and Training (CSEE&T)*, 05 2013, pp. 319–323.

[12] L. Deng, J. Dehlinger, and S. Chakraborty, "Teaching Software Testing with Free and Open Source Software," in *2020 IEEE International Conference on Software Testing, Verification and Validation Workshops (ICSTW)*, Oct. 2020, pp. 412–418. [Online]. Available: https://ieeexplore.ieee.org/document/9155837

[13] SeleniumHQ, "Selenium webdriver: From foundations to framework," https://www.selenium.dev/documentation/en/webdriver/, Open Source Community, 2022.

[14] I. S. Elgrably and S. Ronaldo Bezerra Oliveira, "Model for teaching and training software testing in an agile context," in *2020 IEEE Frontiers in Education Conference (FIE)*, Oct. 2020, pp. 1–9, iSSN: 2377-634X. [Online]. Available: https://ieeexplore.ieee.org/abstract/document/9274117

[15] A. Moubayed, M. Injadat, A. Shami, and H. Lutfiyya, "Relationship Between Student Engagement and Performance in E-Learning Environment Using Association Rules," in *2018 IEEE World Engineering Education Conference (EDUNINE)*, Mar. 2018, pp. 1–6. [Online]. Available: https://ieeexplore.ieee.org/abstract/document/8451005

[16] I. Burman and S. Som, "Predicting students academic performance using support vector machine," in *2019 Amity international conference on artificial intelligence (AICAI)*. IEEE, 2019, pp. 756–759.

[17] E. S. Bhutto, I. F. Siddiqui, Q. A. Arain, and M. Anwar, "Predicting students' academic performance through supervised machine learning," in *2020 International Conference on Information Science and Communication Technology (ICISCT)*, 2020, pp. 1–6.

[18] S. D. A. Bujang, A. Selamat, R. Ibrahim, O. Krejcar, E. Herrera-Viedma, H. Fujita, and N. A. M. Ghani, "Multiclass Prediction Model for Student Grade Prediction Using Machine Learning," *IEEE Access*, vol. 9, pp. 95 608–95 621, 2021, conference Name: IEEE Access. [Online]. Available: https://ieeexplore.ieee.org/document/9468629

[19] M. Khan, S. Naz, Y. Khan, M. Zafar, M. Khan, and G. Pau, "Utilizing machine learning models to predict student performance from lms activity logs," *IEEE Access*, vol. 11, pp. 86 953–86 962, 2023.

[20] Y. Shi, F. Sun, H. Zuo, and F. Peng, "Analysis of learning behavior characteristics and prediction of learning effect for improving college students' information literacy based on machine learning," *IEEE Access*, vol. 11, pp. 50 447–50 461, 2023.

[21] S. Jayasundara, A. Indika, and D. Herath, "Interpretable Student Performance Prediction Using Explainable Boosting Machine for Multi-Class Classification," in *2022 2nd International Conference on Advanced Research in Computing (ICARC)*, Feb. 2022, pp. 391–396. [Online]. Available: https://ieeexplore.ieee.org/abstract/document/9753867

[22] Selenium, "Selenium IDE," https://www.selenium.dev/selenium-ide/, Year.

[23] J. Team, *JUnit 5 User Guide*, JUnit Contributors, 2022. [Online]. Available: https://junit.org/junit5/docs/current/user-guide/

[24] C. Beust and A. Popescu, *TestNG: The Testing Framework for Java*, TestNG Team, 2022. [Online]. Available: https://testng.org/doc/

[25] J. Community, *Jenkins Documentation*, Jenkins Project, 2022. [Online]. Available: https://www.jenkins.io/doc/

[26] Katalon LLC, "Katalon Recorder," https://www.katalon.com/, Year.

[27] F. Pedregosa, G. Varoquaux, A. Gramfort, V. Michel, B. Thirion, O. Grisel, M. Blondel, P. Prettenhofer, R. Weiss, V. Dubourg, J. Vanderplas, A. Passos, D. Cournapeau, M. Brucher, M. Perrot, and E. Duchesnay, "Scikit-learn: Machine learning in Python," *Journal of Machine Learning Research*, vol. 12, pp. 2825–2830, 2011.

[28] R. Alamri and B. Alharbi, "Explainable Student Performance Prediction Models: A Systematic Review," *IEEE Access*, vol. 9, pp. 33 132–33 143, 2021, conference Name: IEEE Access. [Online]. Available: https://ieeexplore.ieee.org/abstract/document/9360749

[29] D. W. Hosmer Jr, S. Lemeshow, and R. X. Sturdivant, *Applied logistic regression*. John Wiley & Sons, 2013, vol. 398.

[30] M. Hearst, S. Dumais, E. Osuna, J. Platt, and B. Scholkopf, "Support vector machines," *IEEE Intelligent Systems and their Applications*, vol. 13, no. 4, pp. 18–28, 1998.

[31] S. Menard, *Applied logistic regression analysis*. Sage, 2002, no. 106.

[32] S. Suthaharan and S. Suthaharan, "Support vector machine," *Machine learning models and algorithms for big data classification: thinking with examples for effective learning*, pp. 207–235, 2016.

[33] B. Charbuty and A. Abdulazeez, "Classification based on decision tree algorithm for machine learning," *Journal of Applied Science and Technology Trends*, vol. 2, no. 01, pp. 20–28, 2021.

[34] L. Breiman, "Random forests," *Machine learning*, vol. 45, pp. 5–32, 2001.

[35] G. Bebis and M. Georgiopoulos, "Feed-forward neural networks," *IEEE Potentials*, vol. 13, no. 4, pp. 27–31, 1994.

[36] S. M. Lundberg and S.-I. Lee, "A unified approach to interpreting model predictions," in *Advances in Neural Information Processing Systems 30*, I. Guyon, U. V. Luxburg, S. Bengio, H. Wallach, R. Fergus, S. Vishwanathan, and R. Garnett, Eds. Curran Associates, Inc., 2017, pp. 4765–4774. [Online]. Available: http://papers.nips.cc/paper/7062-a-unified-approach-to-interpreting-model-predictions.pdf